\documentclass[conference]{IEEEtran}
\IEEEoverridecommandlockouts
\usepackage{cite}
\usepackage{amsmath,amssymb,amsfonts}
\usepackage{algorithmic}
\usepackage{graphicx}
\usepackage{textcomp}
\usepackage{xcolor}
\usepackage{hyperref}

\hypersetup{
    colorlinks=true, 
    linkcolor=black,  
    citecolor=black, 
    filecolor=black, 
    urlcolor=black    
}

\def\BibTeX{{\rm B\kern-.05em{\sc i\kern-.025em b}\kern-.08em
    T\kern-.1667em\lower.7ex\hbox{E}\kern-.125emX}}
\begin{document}

\title{CompressedMediQ:\\ Hybrid Quantum Machine Learning Pipeline for High-Dimensional Neuroimaging Data
\thanks{*Corresponding Author: kuan-cheng.chen17@imperial.ac.uk\\}
}

\author{
\IEEEauthorblockN{
    Kuan-Cheng Chen\IEEEauthorrefmark{2}\IEEEauthorrefmark{3}\IEEEauthorrefmark{1},
    Yi-Tien Li \IEEEauthorrefmark{4}\IEEEauthorrefmark{5},
    Tai-Yu Li \IEEEauthorrefmark{6},
    Chen-Yu Liu\IEEEauthorrefmark{7}\IEEEauthorrefmark{8},
    Po-Heng (Henry) Lee \IEEEauthorrefmark{3}
    Cheng-Yu Chen\IEEEauthorrefmark{4}\IEEEauthorrefmark{9}\IEEEauthorrefmark{10}\IEEEauthorrefmark{11}
}
\IEEEauthorblockA{\IEEEauthorrefmark{2}Department of Electrical and Electronic Engineering, Imperial College London, London, UK}
\IEEEauthorblockA{\IEEEauthorrefmark{3}Centre for Quantum Engineering, Science and Technology (QuEST), Imperial College London, London, UK}

\IEEEauthorblockA{\IEEEauthorrefmark{4}Translational Imaging Research Center, Taipei Medical University Hospital, Taipei, Taiwan}
\IEEEauthorblockA{\IEEEauthorrefmark{5}Research Center for Neuroscience, Taipei Medical University, Taipei, Taiwan}
\IEEEauthorblockA{\IEEEauthorrefmark{5}Ph.D. Program in Medical Neuroscience, College of Medical Science and Technology,\\ Taipei Medical University, Taipei, Taiwan}
\IEEEauthorblockA{\IEEEauthorrefmark{6}National Synchrotron Radiation Research Center, Hsinchu, Taiwan}
\IEEEauthorblockA{\IEEEauthorrefmark{7}Graduate Institute of Applied Physics, National Taiwan University, Taipei, Taiwan}
\IEEEauthorblockA{\IEEEauthorrefmark{8}Hon Hai Research Institute, Taipei, Taiwan}
\IEEEauthorblockA{\IEEEauthorrefmark{9}Department of Radiology, School of Medicine, College of Medicine, Taipei, Taiwan}
\IEEEauthorblockA{\IEEEauthorrefmark{10}Department of Medical Imaging, Taipei Medical University Hospital, Taipei, Taiwan}
\IEEEauthorblockA{\IEEEauthorrefmark{11}Department of Radiology, National Defense Medical Center, Taipei, Taiwan}
}

\maketitle

\begin{abstract}
This paper introduces CompressedMediQ, a novel hybrid quantum-classical machine learning pipeline specifically developed to address the computational challenges associated with high-dimensional multi-class neuroimaging data analysis. Standard neuroimaging datasets, such as large-scale MRI data from the Alzheimer’s Disease Neuroimaging Initiative (ADNI) and Neuroimaging in Frontotemporal Dementia (NIFD), present significant hurdles due to their vast size and complexity. CompressedMediQ integrates classical high-performance computing (HPC) nodes for advanced MRI pre-processing and Convolutional Neural Network (CNN)-PCA-based feature extraction and reduction, addressing the limited-qubit availability for quantum data encoding in the NISQ (Noisy Intermediate-Scale Quantum) era. This is followed by Quantum Support Vector Machine (QSVM) classification. By utilizing quantum kernel methods, the pipeline optimizes feature mapping and classification, enhancing data separability and outperforming traditional neuroimaging analysis techniques. Experimental results highlight the pipeline’s superior accuracy in dementia staging, validating the practical use of quantum machine learning in clinical diagnostics. Despite the limitations of NISQ devices, this proof-of-concept demonstrates the transformative potential of quantum-enhanced learning, paving the way for scalable and precise diagnostic tools in healthcare and signal processing.

\end{abstract}


\section{Introduction}

\begin{figure}[!t]
\centering
\includegraphics[scale=0.35]{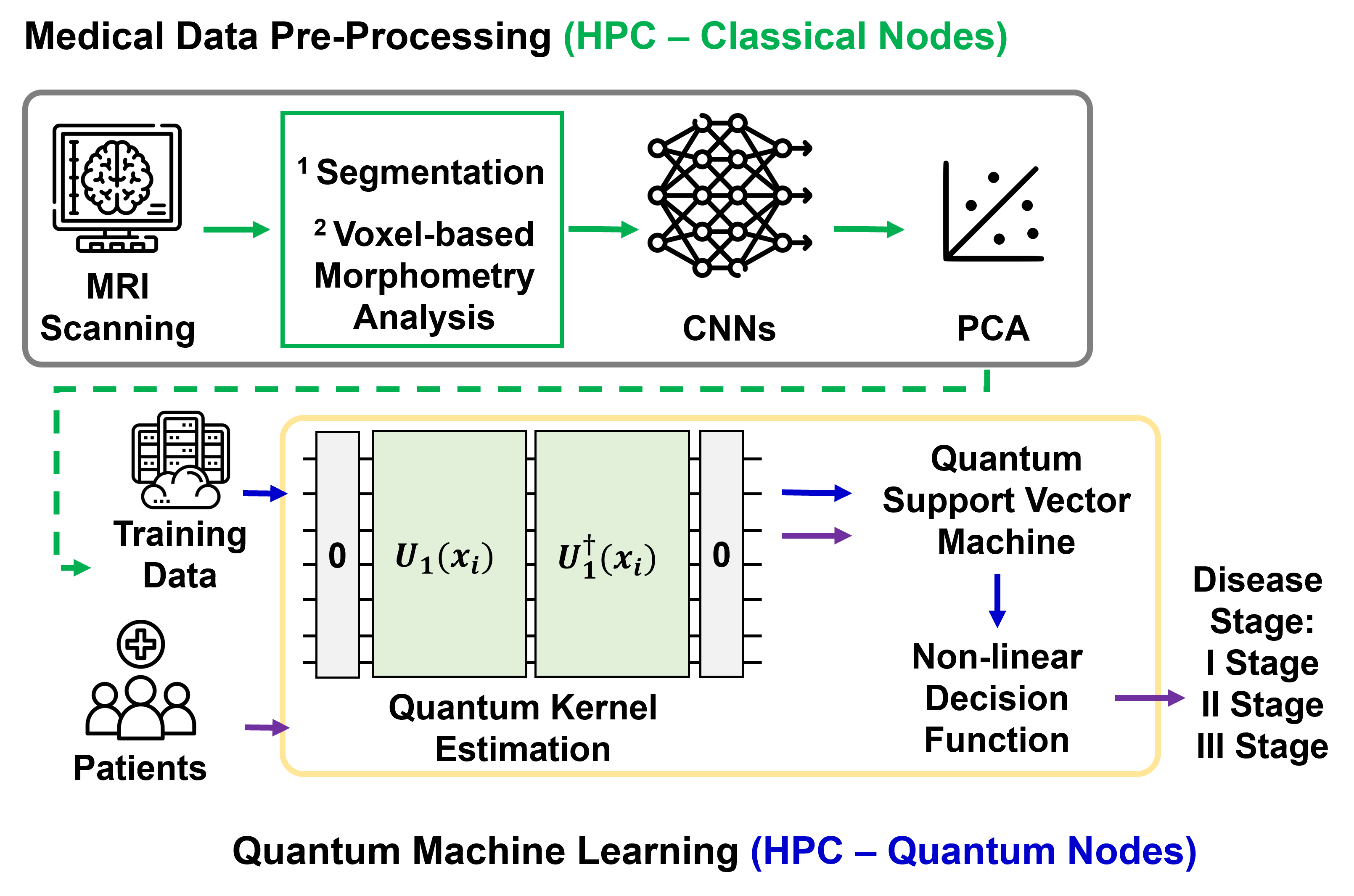}
\caption{Overview of the CompressedMediQ Pipeline for High-Dimensional Neuroimaging Data Analysis. The pipeline integrates classical and quantum computing nodes to process MRI data for disease classification. The workflow begins with data pre-processing on HPC classical nodes, including segmentation and voxel-based morphometry analysis, followed by feature extraction using CNNs and dimensionality reduction via PCA. The extracted features are then input into the quantum machine learning stage, where quantum kernel estimation and a QSVM are employed to perform multi-class classification of disease stages.
 }
\label{fig:concept}
\end{figure}

Magnetic Resonance Imaging (MRI) is a cornerstone of neuroscience, providing critical insights into brain structure and function across a variety of neurological conditions, including Alzheimer's disease (AD)\cite{logothetis2008we, dennis2014functional}. Traditional approaches to MRI analysis have predominantly relied on handcrafted feature extraction methods combined with classifiers such as Support Vector Machines (SVM)\cite{zhang2014classification}. Although these methods offer some success, they often struggle with the high-dimensional and limited sample size nature of MRI data, leading to challenges such as overfitting, prolonged training times, and diminishing accuracy as data dimensionality increases\cite{specht2020current, bandettini2021challenges, beheshti2022classification}. Recently, deep learning models, particularly Convolutional Neural Networks (CNNs), have demonstrated improved performance by automatically extracting high-level features from complex neuroimaging data\cite{wen2020convolutional}. However, these models are not without limitations, often suffering from excessive computational demands, reduced scalability, and sensitivity to noise, making them suboptimal for large-scale, multi-class classification problems typical in MRI applications\cite{beheshti2022classification, sudharsan2023alzheimer}.


This paper introduces CompressedMediQ, a versatile hybrid quantum machine learning (QML) pipeline specifically designed to address the computational challenges associated with high-dimensional MRI data analysis. Standard MRI datasets, often featuring spatial resolutions of about 1 mm and typical dimensions of 256 x 256 x 150 voxels, result in approximately 10 million individual voxels per scan. These large data sizes, ranging from several hundred megabytes to several gigabytes per scan session, pose significant challenges for traditional machine learning approaches in terms of computational efficiency and accuracy\cite{makkie2018distributed}. While quantum machine learning has shown superior performance in various applications\cite{chen2024quantum2, lin2024quantum, ho2024quantum, liu2024qtrl, li2022BIBE}, the limited qubit availability for data encoding in the NISQ (Noisy Intermediate-Scale Quantum) era remains a major constraint\cite{li2022concentration,bharti2022noisy}. CompressedMediQ addresses these issues by integrating quantum computing with classical ML techniques, enhancing processing speed and accuracy\cite{huang2021power, caro2022generalization}. The pipeline leverages quantum properties such as superposition and entanglement to improve the efficiency of handling massive datasets. As depicted in Fig. \ref{fig:concept}, CompressedMediQ consists of two primary stages: (1) MRI pre-processing using classical high-performance computing (HPC) nodes, and (2) a hybrid quantum-classical workflow that incorporates Quantum Support Vector Machines (QSVM) with quantum feature mapping to encode classical data into quantum states. This approach enables the exploration of high-dimensional feature spaces with enhanced separability between classes, effectively overcoming the limitations of traditional methods\cite{chen2024quantum, chen2024cutn}.

In this paper, we detail the architecture of the CompressedMediQ pipeline, outline the specific quantum algorithms used, and present comparative results highlighting its advantages. Our findings underscore the potential of hybrid QML frameworks in overcoming the scalability and accuracy limitations of classical methods, offering a new paradigm for large-scale neuroimaging data analysis. By demonstrating a proof-of-concept application in dementia classification—a particularly challenging problem in neuroimaging research\cite{tartaglia2011neuroimaging}—this work validates the practical benefits of integrating quantum-enhanced learning techniques. CompressedMediQ not only advances the field of MRI-based AD classification but also sets the foundation for future research in applying quantum-enhanced algorithms to complex medical data challenges, paving the way for scalable and accurate diagnostic solutions\cite{sudharsan2023alzheimer}.

\section{Methodology}

\subsection{MRI Data Collection}
The data used in this study were sourced from two primary databases: the Alzheimer’s Disease Neuroimaging Initiative (ADNI)\cite{petersen2010alzheimer} and Neuroimaging in Frontotemporal Dementia (NIFD)\cite{rohrer2013neuroimaging}. The subjects were categorized into six distinct groups based on their CDR-SB scores\cite{o2008staging}, which served as the criteria for classification. These groups included normal controls, with CDR-SB scores of 0; individuals with questionable cognitive impairment, whose CDR-SB scores ranged from 0.5 to 2.5; subjects classified with very mild dementia, indicated by CDR-SB scores between 3 and 4; those with mild dementia, identified by CDR-SB scores from 4.5 to 9; individuals with moderate dementia, whose scores ranged between 9.5 and 15.5; and finally, subjects with severe dementia, characterized by CDR-SB scores ranging from 16 to 18. This stratification provided a structured framework for analyzing varying stages of cognitive decline, supporting the study’s goal of exploring neuroimaging data across a spectrum of dementia severities.


\subsection{MRI Data Pre-Processing}


Prior to inputting the data into our specialized auto-segmentation neural network models, the MRI scans underwent a rigorous preprocessing pipeline designed to optimize data quality and reliability. This preprocessing approach was essential for standardizing the structural MRI data, correcting potential artifacts, and accurately isolating relevant brain tissues for further analysis. The following subsections provide a detailed account of each step in this comprehensive preprocessing protocol.

\begin{figure}[!b]
\centering
\includegraphics[scale=0.34]{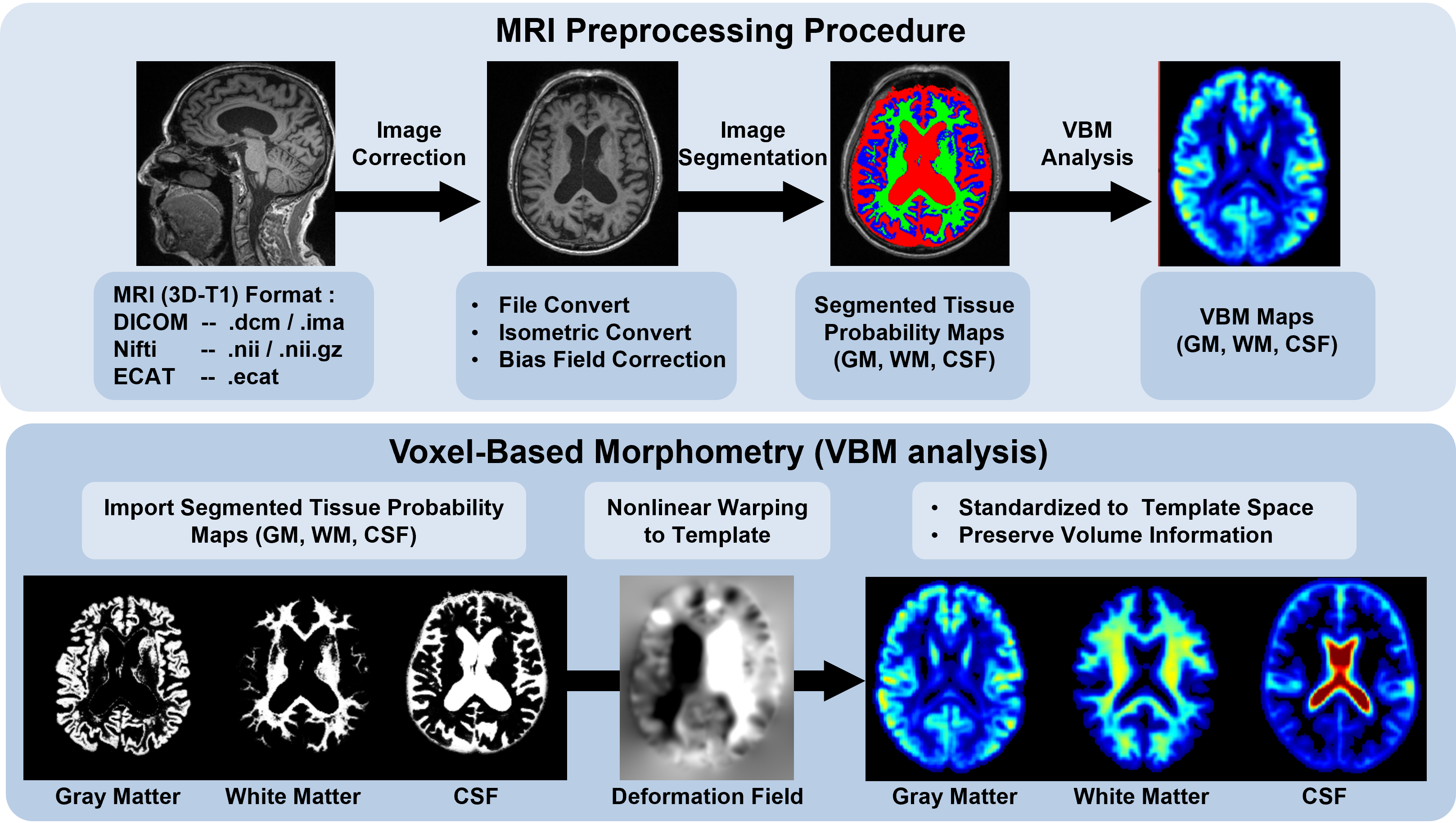}
\caption{Overview of the MRI Preprocessing Procedure and VBM Analysis, illustrating the steps from image correction and segmentation to VBM map generation, and nonlinear warping of segmented tissue probability maps to standardized template space for preserving volumetric information of GM, WM, and CSF.
 }
\label{fig:preprocess}
\end{figure}

\begin{figure*}[!t]
\centering
\includegraphics[scale=0.35]{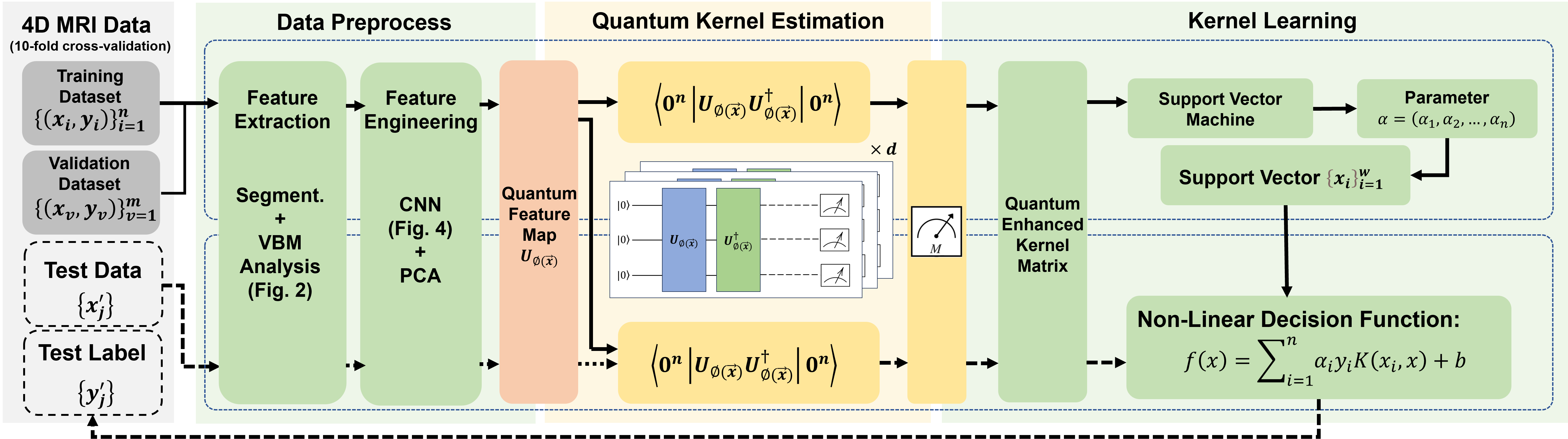}
\caption{Overview of the CompressedMediQ pipeline illustrating the hybrid quantum-classical workflow, including data preprocessing with CNN and PCA, quantum kernel estimation, and kernel learning using QSVM for enhanced multi-class classification of high-dimensional MRI data.
}
\label{fig:scheme}
\end{figure*}

\subsubsection{Brain tissue segmentation}
The 3D T1-weighted structural MRI scans were acquired from all participants and underwent a comprehensive preprocessing pipeline using the unified segmentation algorithm available in the SPM12 software package\cite{ashburner2014spm12}. SPM12 is a MATLAB-based toolset designed for advanced statistical analysis of neuroimaging data. This preprocessing phase is crucial for standardizing the MRI scans, ensuring that subsequent analyses are both accurate and reliable.

The segmentation process began with bias field correction, which compensates for intensity inhomogeneities caused by magnetic field distortions during MRI acquisition. Following this, the algorithm performed tissue classification to segment the brain into distinct tissue classes: gray matter (GM), white matter (WM), and cerebrospinal fluid (CSF). Additionally, probability maps were generated for non-brain tissues, such as the skull and scalp. The segmentation utilized medium bias regularization and incorporated a reference brain probability map provided by SPM12, which enhances the precision and consistency of the tissue classification. This rigorous approach ensures high-quality data that is essential for subsequent statistical analyses.

\subsubsection{Voxel-based morphometry (VBM) analysis}
Following the initial segmentation, the MRI data underwent further refinement using the Diffeomorphic Anatomical Registration Through Exponentiated Lie Algebra (DARTEL)-based spatial normalization technique\cite{chen2012mapping}, implemented via SPM12 software. This advanced normalization method is particularly advantageous for studies involving multi-subject comparisons, as it aligns the structural brain images into a common anatomical framework. The precise alignment facilitated by DARTEL helps in reducing inter-subject variability, ensuring that each image conforms closely to a standard template, thereby enhancing the robustness of subsequent analyses.

During the normalization process, certain brain regions may expand or contract to match the template’s anatomical structure. To retain the original volumetric characteristics of these regions, a modulation step is employed, which scales each voxel in the segmented images by the Jacobian determinant derived from the spatial transformation. This modulation step preserves the integrity of tissue volume information (e.g., GM, WM, CSF) by ensuring that the overall signal remains consistent before and after spatial normalization. This meticulous preprocessing approach is designed to minimize variability from individual anatomical differences, thus enabling more precise and reliable statistical analyses of the structural MRI data.

\subsection{Hybrid Classic-Quantum Network Architecture}

The CNN-PCA-QSVM architecture proposed in this study combines classical CNNs with QSVMs to enhance the classification of neuroimaging data, as illustrated in Figure 1. The CNN component of the architecture accepts modulated GM and CSF maps as inputs, derived from MRI data to extract essential features relevant to dementia classification. The CNN comprises two primary sections: a down-convolutional section for reducing spatial dimensions and extracting key features, and a fully connected section designed to model the non-linear relationships among these features. Each basic convolutional module within the down-convolutional section starts with a 3x3x3 convolutional layer followed by batch normalization, with shortcut connections merging the input layers and outputs of the stacked convolutional layers to facilitate identity mapping and mitigate the vanishing gradient problem. The Scaled Exponential Linear Unit (SELU) is employed as the activation function, followed by a 2x2x2 max pooling operation with a stride of 2 for down-sampling. These convolutional modules are replicated four times throughout the architecture, with the 1st, 3rd, and 5th modules augmented by a Convolutional Block Attention Module (CBAM) to refine features adaptively before max pooling.

\begin{figure}[!b]
\centering
\includegraphics[scale=0.45]{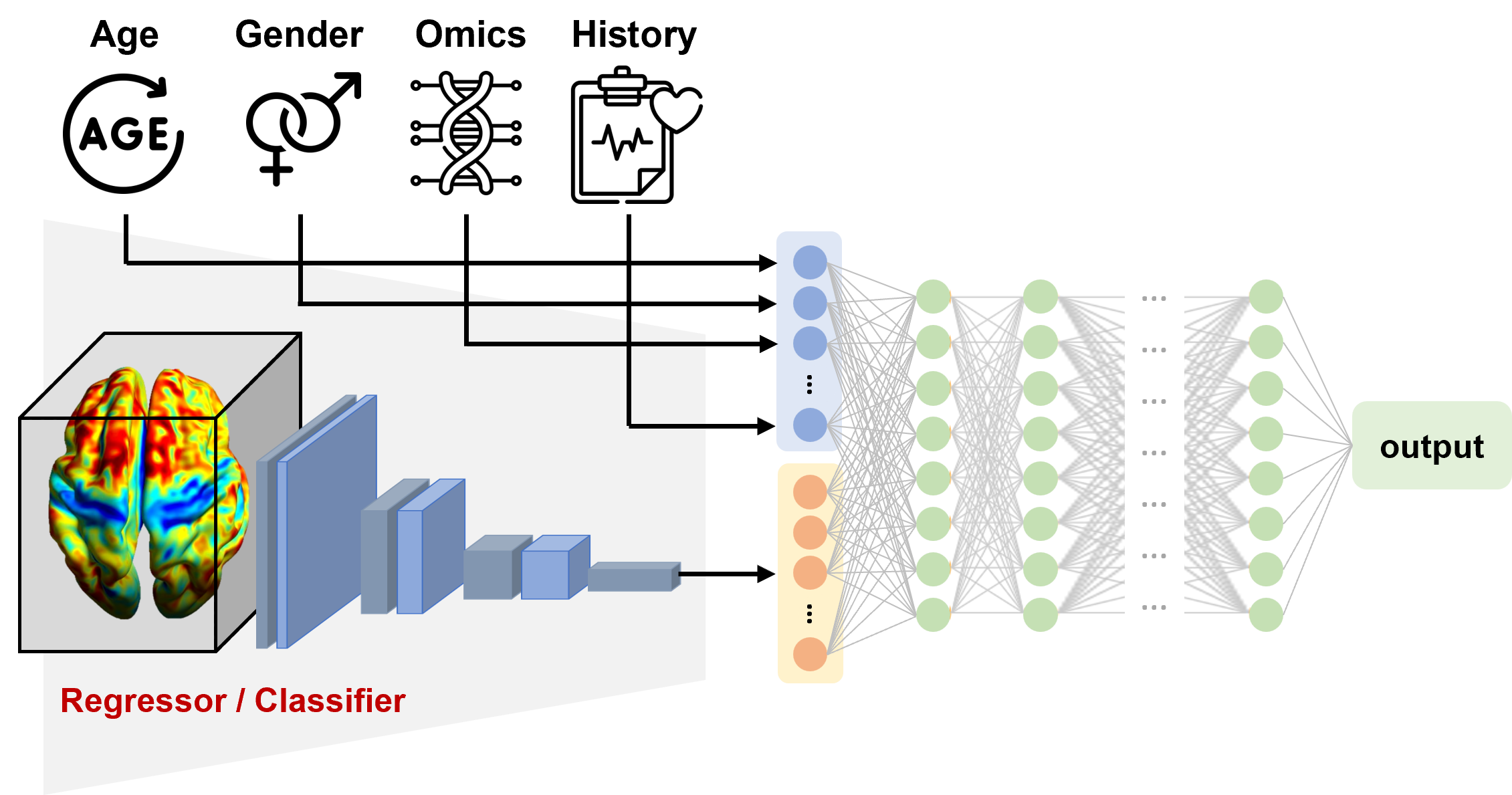}
\caption{The hybrid framework integrates MRI-derived features with demographic, genomic, and clinical history inputs into a combined regressor/classifier model for enhanced neuroimaging analysis and prediction.
 }
\label{fig:cnn}
\end{figure}

Following the convolutional layers, a dedicated convolution module consolidates the feature maps, which are then flattened into a one-dimensional array and connected to four fully connected layers, each containing 32 nodes. Demographic data, including subjects’ age and sex, are integrated with the feature nodes after the first fully connected layer to enrich the feature space. To reduce the high-dimensional output from the CNN, Principal Component Analysis (PCA) is utilized to condense the data into a reduced feature space with a dimension of 8, crucial for optimizing the subsequent quantum encoding phase. This PCA step not only compresses the data into a manageable size but also enhances computational efficiency, particularly in the NISQ era.

After dimensionality reduction, the data are encoded into both classical and quantum feature spaces. Quantum kernel methods complement the CNN-based feature extraction by leveraging advanced quantum computing techniques for classification. SVMs are particularly effective in high-dimensional, non-linear classification tasks, extending their capabilities through kernel functions that project data into higher-dimensional feature spaces. QSVMs extend this concept further by mapping classical data into quantum states within the Hilbert space using quantum feature maps\cite{schuld2019quantum}. This process involves rotational entanglement operations:

\begin{equation}
U_{\Phi}(\mathbf{x}) = \exp\left(i \sum_{S \subseteq [n]} \phi_S(\mathbf{x}) \prod_{k \in S} Z_k \right),
\end{equation}

which encode the classical features into quantum states, enabling the formation of quantum kernels that capture complex, high-dimensional relationships with enhanced computational efficiency. The quantum kernel between two data points is calculated as:

\begin{align}
K(\mathbf{x}_i, \mathbf{x}_j) &= |\langle \psi(\mathbf{x}_i) | \psi(\mathbf{x}_j) \rangle|^2 \\
&= |\langle 0^{\otimes N} | U^\dagger(\mathbf{x}_i) U(\mathbf{x}_j) | 0^{\otimes N} \rangle|^2,
\end{align}

utilizing quantum parallelism to achieve significant computational speedups, often scaling as \( O(\log(n)) \) compared to classical methods for kernel calculations involving \( n^2 \) data point pairs \cite{gentinetta2024complexity}. This hybrid CNN-QSVM approach effectively combines deep learning's feature extraction capabilities with quantum-enhanced kernel methods, pushing the boundaries of data dimensionality handling and computational efficiency in neuroimaging classification tasks.

\section{Result}
\begin{figure}[!t]
\centering
\includegraphics[scale=0.41]{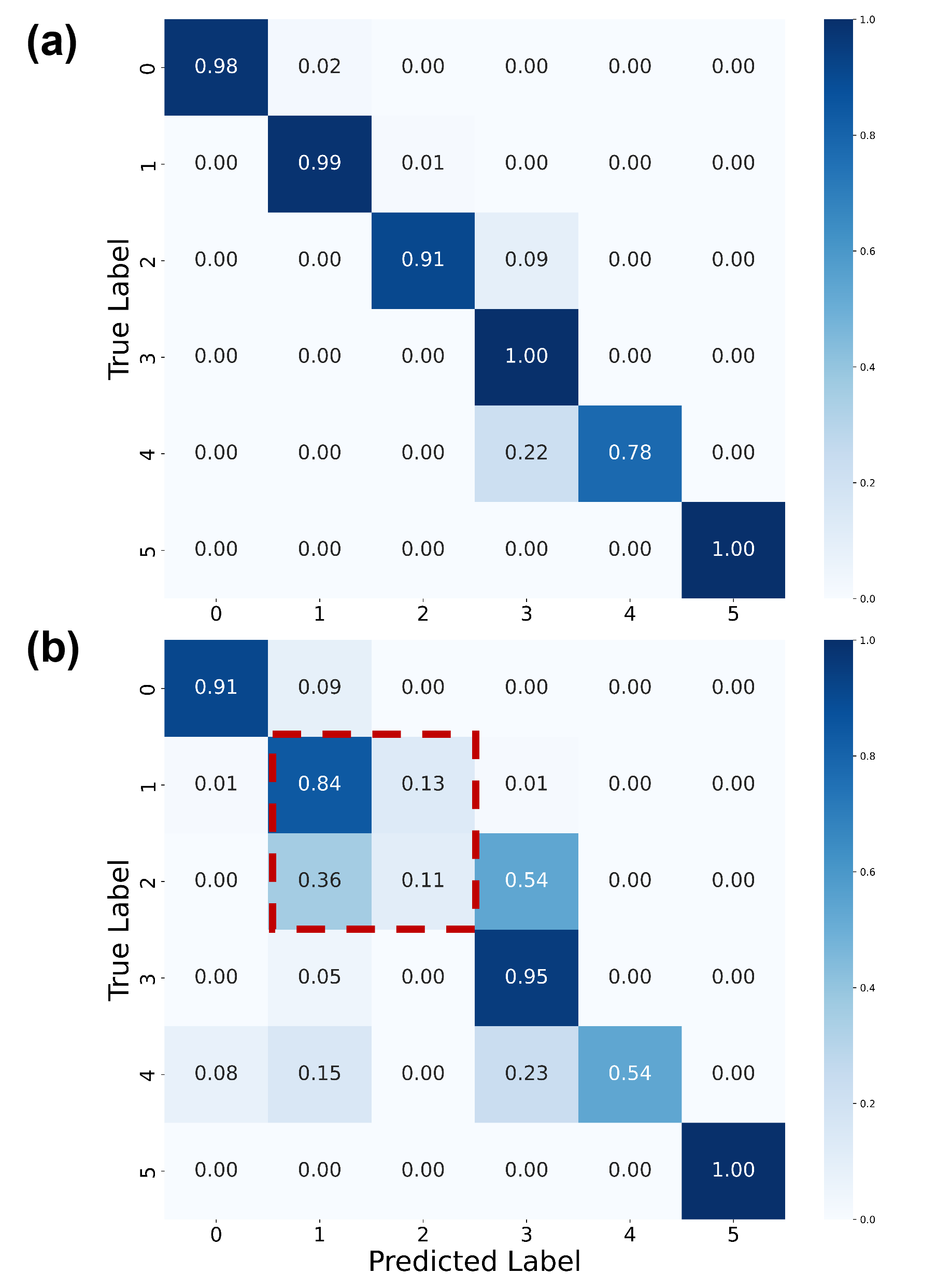}
\caption{Confusion matrices comparing Quantum (a) and Classical SVM (b) models, highlighting superior accuracy and reduced misclassification in the Quantum model (96.1\%) versus the Classical SVM model (78.8\%), particularly within the marked misclassification region.
 }
\label{fig:result}
\end{figure}

The evaluation of the hybrid CNN-QSVM pipeline, as illustrated in Fig. \ref{fig:result}, highlights the superior performance of the quantum model compared to the classical approach in dementia classification. The quantum model achieved high classification accuracy, particularly in distinguishing early stages of cognitive decline, such as normal controls, questionable cognitive impairment, and very mild dementia, with precision rates of 98\%, 99\%, and 91\%, respectively. It maintained strong performance across all categories, with minimal misclassification between mild, moderate, and severe dementia stages. These results demonstrate the model’s ability to effectively manage complex neuroimaging data, providing precise classification across a broad spectrum of dementia severities.

In contrast, the classical SVM model, also depicted in Fig. \ref{fig:result}, struggled significantly, particularly in distinguishing mild and moderate dementia, as evident from the confusion matrix. The model frequently misclassified cases within these stages, with up to 36\% of instances confused between mild and moderate dementia, highlighted within the marked region of the matrix. This pattern reflects the limitations of the classical approach in capturing subtle differences in intermediate stages of cognitive impairment. The comparison underscores the enhanced accuracy and reliability of the quantum model, highlighting its potential as a valuable tool for neuroimaging-based diagnostics in dementia. These findings validate the integration of QML within clinical pipelines, offering a significant advancement in diagnostic precision for neurodegenerative diseases.

\section{Conclusion}
This study introduces CompressedMediQ, a hybrid pipeline that integrates classical and quantum computing methods to address the challenges associated with high-dimensional neuroimaging data. By leveraging advanced MRI pre-processing, CNN-based feature extraction/reduction, and quantum-enhanced classification through QSVM, the proposed pipeline demonstrates substantial improvements in accuracy and computational efficiency over traditional models. The results highlight the quantum model's superior ability to accurately differentiate between various stages of dementia, especially in cases where classical approaches struggle to capture nuanced patterns of cognitive decline. Despite current hardware limitations inherent in NISQ devices, this proof-of-concept validates the viability of quantum-enhanced machine learning frameworks in real-world applications such as healthcare and signal processing\cite{kourou2015machine,uddin2019comparing}. 

The study’s findings suggest that the hybrid approach not only addresses the computational bottlenecks of traditional methods but also offers a scalable solution capable of advancing clinical diagnostics. Future work will focus on expanding the quantum-classical pipeline with techniques such as quantum error mitigation\cite{strikis2021learning,chen2023short}, distributed quantum computing\cite{burt2024generalised}, and resource management\cite{chen2024noise} to accommodate larger datasets and explore its applicability to other complex neuroimaging tasks within quantum HPC systems\cite{humble2021quantum}.



\bibliographystyle{siamurl}
\bibliography{references}

\end{document}